\documentclass[12pt,a4paper]{article}
\usepackage[utf8]{inputenc}
\usepackage{amsfonts}
\usepackage{amsmath,amssymb,graphicx,epsfig}
\usepackage{enumerate,amsthm,epstopdf}
\usepackage{bbm}
\usepackage{dsfont}
\usepackage{algpseudocode}
\usepackage{algorithmicx}
\usepackage{algorithm}
\usepackage{listings}
\usepackage{color}
\usepackage{hyperref}
\usepackage[makeroom]{cancel}
\usepackage{verbatim}

\def\bP{{\mathbb P}}

\def\f0{{\mathbf 0}}

\theoremstyle{definition}

\usepackage[right=1.3in, left=1.3in, bottom=1in]{geometry}

\title{A probabilistic interpretation of replicator-mutator dynamics}
\author{\"Omer Deniz Aky{\i}ld{\i}z\thanks{The author is with the Department of Signal Theory and Communications, Carlos III University of Madrid. Email: \texttt{oakyildi@pa.uc3m.es} and web: \texttt{http://akyildiz.me}.}}
  
\begin{document}
\maketitle
\begin{abstract}
In this note, we investigate the relationship between probabilistic updating mechanisms and discrete-time replicator-mutator dynamics. We consider the recently shown connection between Bayesian updating and replicator dynamics and extend it to the replicator-mutator dynamics by considering prediction and filtering recursions in hidden Markov models (HMM). We show that it is possible to understand the evolution of the frequency vector of a population under the replicator-mutator equation as a posterior predictive inference procedure in an HMM. This view enables us to derive a natural dual version of the replicator-mutator equation, which corresponds to updating the filtering distribution. Finally, we conclude with the implications of the interpretation and with some comments related to the recent discussions about evolution and learning.
\end{abstract}

\section{Introduction}
Recently, it has been shown that there is a connection between Bayesian updating and replicator dynamics \cite{shalizi2009dynamics,harper2009replicator}. If the frequency distribution of a population at a given time is interpreted as a probability vector, then evolutionary dynamics models correspond to updating these probabilities. This, in turn, can be seen as updating probability distributions over time via Bayesian updating as shown in \cite{shalizi2009dynamics}. In this note, we first review the proposed relationship in detail to set up the context. Then, we briefly extend the previous results to the replicator-mutator case, where in addition to the replication, there is mutation dynamics. Then we give a filtering recursion which we call as \textit{the mutator-replicator equation}. Finally, we discuss some of the implications about seeing evolutionary processes as probabilistic updating mechanisms.
\\ \\
\textbf{Notation.} Throughout, $x_t^i$ will denote the frequency of type $i$ at time $t$ and $x_t$ will denote the whole distribution (a vector, in this case) at time $t$. The fitness function for type $i$ is denoted with $f_i$. The average fitness will be given by $\bar{f}_t = \sum_{j} x_t^j f_j$. We will denote a generic probability measure over a discrete state space with $\bP_t$ to denote the distribution of the population at time $t$ and $\varphi_t(\cdot)$ will be the likelihood. We assume there are $n$ competing types. We define a state-space $\mathcal{S} = \{1,\ldots,n\}$. To highlight the correspondence, we note that $x_t^i = \bP_t(\theta = i)$ and $\varphi_t(i) = f_i(\cdot)$ (the argument will depend on the context). Relevant notation will be introduced further when needed.
\section{Evolutionary dynamics as Bayesian inference}
\subsection{Replicator dynamics as Bayesian updating}
Consider the \textit{discrete time} replicator dynamics which is defined as \cite{hofbauer1998evolutionary},
\begin{align}\label{eqRepDyn}
x_t^i = x_{t-1}^i \frac{f_i(x_{t-1})}{\bar{f}_t}
\end{align}
where $x_t^i$ is the frequency of the population of $i$th type and $f_i$ is the fitness function of $i$th type and $\bar{f}_t$ is the mean fitness given by,
\begin{align*}
\bar{f}_t = \sum_j x_{t-1}^j f_j(x_{t-1}).
\end{align*}
Next, let us define a random variable $\theta$ defined on $\mathcal{S}$ where $\mathcal{S} = \{1,\ldots,n\}$. This random variable models the probability of a single individual belongs to a certain type, as the frequency can be interpreted this way \cite{kolmogorov1950foundations}. Therefore, we write $x_t^i = \bP_t(\theta = i)$\footnote{Note that, with a slight abuse of notation, $\bP(i)$ and $\bP(\theta = i)$ denotes the same quantity for each $i \in \mathcal{S}$.}.

Assume that, at time $t-1$, we have $\bP_{t-1}(\theta)$ describing the frequencies of species. Given the likelihood (or the fitness potential) $\varphi_t(\theta)$, we can update this probability distribution via the Bayes' rule as \cite{shalizi2009dynamics},
\begin{align}\label{eqBayesRule}
\bP_t(\theta) = \bP_{t-1}(\theta) \frac{\varphi_t(\theta)}{\sum_{\theta' \in \mathcal{S}} \bP_{t-1}(\theta') \varphi_t(\theta')}.
\end{align}
It is possible to see that \eqref{eqRepDyn} and \eqref{eqBayesRule} describe the exact same relationship \cite{shalizi2009dynamics,harper2009replicator}. For this to work, we need to put $f_i(x_{t-1}) = \varphi_t(i)$, i.e., define a likelihood that depends on the whole distribution of previous time. This interpretation of the replicator dynamics as Bayesian updating was pointed out by \cite{shalizi2009dynamics,harper2009replicator}. The replicator equation is more general since the likelihood in the Bayesian context does not depend on the whole distribution \cite{shalizi2009dynamics}.
\subsection{Replicator-mutator dynamics as Bayesian inference}
If the replicator dynamics is Bayesian updating, it is natural to expect that there must be a dynamic Bayesian version for the replicator-mutator dynamics. As a straightforward dynamic extension of Bayesian updating, it is tempting to consider the prediction and filtering recursions (the latter is known as \textit{the optimal Bayesian filter} \cite{anderson1979optimal}, see \cite{sarkka2013bayesian} for an accessible treatment) to see what they mean in the evolutionary dynamics context.

To begin with, we recall that the replicator-mutator equation, which has received a significant attention and widely used, can be described as \cite{page2002unifying, harper2012stability},
\begin{align}\label{eqQuasi}
x_t^i = \frac{\sum_j f_j(x_{t-1}) x_{t-1}^j K_{ji}}{\sum_{j} f_j x_{t-1}^j}
\end{align}
where $K_{ji}$ is a transition matrix, i.e.,
\begin{align*}
\sum_{i} K_{ji} = 1
\end{align*}
for every $j \in \mathcal{S}$. Now consider a Markov chain $(\theta_t)_{t\geq 0}$ with a transition matrix,
\begin{align*}
K(\theta_t = i | \theta_{t-1} = j) = K_{ji}.
\end{align*}
Our aim is to come up with a probabilistic interpretation of \eqref{eqQuasi} as an update of conditional distributions of $(\theta_t)_{t\geq 0}$ given the sequence of ``observations''.

We derive the replicator-mutator equation as a probabilistic update by putting $\varphi_{t-1}(\theta_{t-1} = i) = f_i(x_{t-1})$. Then in the probabilistic setup, we get,
\begin{align}\label{eqPredRecur}
{\bP}_t(\theta_t) = \frac{\sum_{\theta'_{t-1}\in\mathcal{S}} {\bP}_{t-1}(\theta_{t-1}') K(\theta_{t} | \theta_{t-1}') \varphi_{t-1}(\theta_{t-1}')}{\sum_{\theta'_{t-1} \in\mathcal{S}} {\bP}_{t-1}(\theta_{t-1}') \varphi_t(\theta_{t-1}')}.
\end{align}
We can immediately recognize this recursion as the prediction recursion of a hidden Markov model (HMM)\footnote{This might be easier to see from the recursion over a continuous space $\Theta$ and with an explicit observation sequence $y_{1:t}$. Consider the likelihood $g(y_t|\theta_t)$ and the transition density $K(\theta_t|\theta_{t-1})$. Then, the following holds,
\begin{align*}
p(\theta_t|y_{1:t-1}) = \frac{\int_{\Theta} p(\theta_{t-1}|y_{1:t-2}) K(\theta_t|\theta_{t-1}) g(y_{t-1}|\theta_{t-1}) \mbox{d}\theta_{t-1}}{\int_\Theta p(\theta_{t-1}|y_{1:t-2}) g(y_{t-1}|\theta_{t-1})\mbox{d}\theta_{t-1}}
\end{align*}
as the density in the nominator can be written as $p(\theta_t,y_{t-1}|y_{1:t-2})$ and the density in the denominator can be written as $p(y_{t-1}|y_{1:t-2})$. Note that, for this to hold $g(y_{t-1}|\theta_{t-1})$ need not to be a probability density.

See the concluding sections for the meaning of conditioning on $y_{1:t-1}$ in this context.} \cite{van2008hidden}. That is, $\bP_{t-1}(\theta_{t-1})$ is the predictive distribution at time $t-1$ given all the data up to time $t-2$. The recursion \eqref{eqPredRecur} is a map acting between the space of probability distributions and it maps the predictive distribution $\bP_{t-1}(\theta_{t-1})$ of time $t-1$ to the predictive distribution $\bP_t(\theta_t)$ of time $t$. So the replicator-mutator dynamics can be thought of as employing Bayesian prediction in an HMM with a likelihood depends on the whole probability distribution of the previous time. \textit{Observations} in this setting are implicit in the fitness functions as the values of the fitness functions can be reinterpreted as \textit{evaluations} of the likelihood with an implicit data sequence. The posterior predictive distribution over hidden states exactly coincides with the frequencies of the population given by the replicator-mutator equation.
\subsubsection*{From replicator-mutator to mutator-replicator}
Now as an obvious next step, we can investigate the filtering recursion. To keep the notation similar, let us denote the filtering distribution with $\tilde{\bP}_t$ and corresponding population vector with $\tilde{x}_t$. Then the filtering recursion can be written as,
\begin{align}\label{eqFiltRecur}
\tilde{\bP}_t(\theta_t) = \frac{\varphi_t(\theta_t) \sum_{\theta'_{t-1}\in\mathcal{S}} \tilde{\bP}_{t-1}(\theta_{t-1}') K(\theta_{t} | \theta_{t-1}')}{\sum_{\theta_t' \in \mathcal{S}} \varphi_t(\theta_t') \sum_{\theta'_{t-1} \in\mathcal{S}} \tilde{\bP}_{t-1}(\theta_{t-1}') K(\theta_{t}'| \theta_{t-1}')}.
\end{align}
In terms of the relevant literature, we can rewrite the filtering recursion as,
\begin{align}\label{eqQuasiNew}
\tilde{x}_t^i = \frac{f_i({x}_t) \sum_j \tilde{x}_{t-1}^j K_{ji}}{\sum_i f_i({x}_t) \sum_j \tilde{x}_{t-1}^j K_{ji}},
\end{align}
where $x_t$ is defined component-wise with ${x}^i_t = \sum_j \tilde{x}_{t-1}^j K_{ji}$ (which is actually the predictive distribution in the Bayesian sense, as the notation suggests -- see above). This equation is different from the replicator-mutator equation. We refer to it as \textit{the mutator-replicator equation} and it has a natural interpretation related to the replicator-mutator equation. We have shown that the Eq.~\eqref{eqQuasi} can be interpreted as a prediction recursion (that is the most recent distribution of the population after the last replication-mutation steps, but \textit{before} the next replication step). Similarly, the Eq.~\eqref{eqQuasiNew} can be interpreted as the distribution of the population after the last mutation-replication steps but \textit{before} the next mutation step, hence the name \textit{mutator-replicator}. These two equations are complementary to each other in a very natural sense and it would be interesting to see if the latter recursion could be useful in the study of evolutionary dynamics.
\section{Discussion}
The interaction between two fields can be productive if this view can be taken to its natural conclusion. In the evolutionary dynamics context, there are several versions of the replicator-mutator equations which is a rich family of tools, aiming at modeling different mechanisms. One can try to make sense of some of those from a probabilistic modeling perspective to uncover the underlying probabilistic structure. Reversely, tools of the computational Bayesian statistics can be used to analyze these models, by taking an inference view on the problem or transferring the already well-known theoretical results to understand the dynamics of the models in evolutionary dynamics from a different perspective. As an example, if we compute $\sum_t \bar{f}_t$, this coincides with the marginal likelihood in Bayesian computation, which is also called as \textit{the model evidence}. This is a quantity which enables us to rank different models. It could be fruitful to think about its applications in the evolutionary dynamics context, e.g., on whether it can be used to test different mutation mechanisms against each other given a specific fitness landscape.

We remark that this short note only proposes that \textit{some models of evolution} can be understood as dynamic Bayesian updating mechanisms. This is also related to recent discussions about \textit{how evolution can learn}, see e.g. \cite{watson2016can}. While it is true that, under this particular replicator-mutator (or mutator-replicator) model, the distribution of the population is conditioned on all the previous evaluations (that can be regarded as implicit observations from the environment), it does not follow immediately that the evolution can utilize past information entirely, as a learning algorithm would. Given the dynamic view here, one can argue that evolution works more like a \textit{tracking} algorithm, rather than a learning one\footnote{By \textit{learning}, we mean parameter estimation (or fitting) in a statistical model (either maximum-likelihood or Bayesian).}. For many practical dynamic models, the posterior probability distribution (the filter) tends to \textit{forget} the past exponentially fast under mild conditions \cite{Cappe2005}, which means that only the most recent environment evaluations, the recent state of the system, and the recent mutations might be relevant rather than the entire past. Although there is an abstract possibility that the mutation mechanisms can evolve themselves, it still does not imply that evolution can learn beyond adaptation to the current structure of the environment. Mathematically, if we can define a mutation matrix $K_{\gamma_t}(\theta_t|\theta_{t-1})$ with a parameter vector $\gamma_t$, which parameterizes the mutation mechanism, by generating variants of $\gamma_t$ and a nested structure (running slightly different mutation mechanisms for each subgroup), the parameter vector $\gamma_t$ can be adapted to the environment (e.g. see \cite{crisan2013nested} for such an algorithm for continuous state-space models within the Monte Carlo framework). From a statistical perspective, it would correspond to adapting the parameters of the transition model of an HMM\footnote{There is some evidence that such a mechanism exists \cite{hull2017environmental}}. However, even if there is such a mechanism, it can still be regarded as some form of \textit{tracking} (as real-time adaptation) rather than \textit{learning} as we know it.
\section{Conclusions}
We believe that a probabilistic view of the replicator-mutator equation can help unifying well-known ideas from probabilistic modeling and evolutionary dynamics, which then can help merging the well-known tools of computational study of two separate lines of research. The new tools emerging from this relationship can lead to a number of fruitful ideas and help the development of more elaborate models of evolutionary phenomena.
\newpage
\bibliographystyle{unsrt}
\bibliography{../../../CommonLatexFiles/draft} 
\end{document}